\documentclass{elsart}

\usepackage{amsmath}
\usepackage{amssymb}
\usepackage{graphicx}

\newcommand {\ignore}[1]{}

\def\lsim{\mathrel{\rlap{\lower4pt\hbox{\hskip1pt$\sim$}}
    \raise1pt\hbox{$<$}}}         
\def\gsim{\mathrel{\rlap{\lower4pt\hbox{\hskip1pt$\sim$}}
    \raise1pt\hbox{$>$}}}         

\begin{document}

\raisebox{8mm}[0pt][0pt]{\hspace{12cm}\vbox{hep-ph/yymmdd\\IFIC/01-44}}

\begin{frontmatter}


\title{A non-resonant dark-side solution to the solar neutrino problem}

\author{O. G. Miranda$^2$},
\ead{Omar.Miranda@fis.cinvestav.mx}
\author{C. Pe{\~n}a-Garay$^1$},
\ead{penya@hal.ific.uv.es}
\author{T. I. Rashba$^3$},
\ead{rashba@izmiran.rssi.ru}
\author{V. B. Semikoz$^3$} and
\ead{semikoz@orc.ru, semikoz@ific.uv.es}
\author{J.~W.~F. Valle$^1$}
\ead{valle@ific.uv.es}

\address{$^1$Instituto de F\'{\i}sica Corpuscular -- C.S.I.C.,
  Universitat de Val{\`e}ncia \\
  Edificio Institutos, Apt.\ 22085, E--46071 Val{\`e}ncia, Spain}

\address{$^2$Departamento de F\'{\i}sica, Centro de Investigaci{\'o}n y de
  Estudios Avanzados Apdo. Postal 14-740 07000 Mexico, DF, Mexico}

  \address{$^3$ Institute of Terrestrial Magnetism,
    Ionosphere and Radio Wave Propagation of the Russian Academy of Sciencies,
    142190, Troitsk, Moscow region, Russia}

\begin{abstract}
  We re-analyse spin-flavour precession solutions to the solar
  neutrino problem in the light of the recent SNO CC result as well as
  the 1258--day Super-Kamiokande data and the upper limit on solar
  anti-neutrinos. In a self-consistent magneto-hydrodynamics approach
  the resulting scheme has only 3 effective parameters: $\Delta m^2$,
  $\mu B_\perp$ and the neutrino mixing angle $\theta$.  We show how a
  rates-only analysis for fixed $\mu B_\perp$ slightly favours
  spin-flavour precession (SFP) solutions over oscillations (OSC).  In
  addition to the resonant solution (RSFP for short), there is a new
  non-resonant solution (NRSFP) in the ``dark-side''.  Both RSFP and
  NRSFP lead to flat recoil energy spectra in excellent agreement with
  the latest SuperKamiokande data.  We also show that in the presence
  of a neutrino transition magnetic moment of $10^{-11}$ Bohr
  magneton, a magnetic field of 80 KGauss eliminates all large mixing
  solutions other than the so-called LMA solution.

\end{abstract}

\end{frontmatter}

\section{Introduction}
\label{sec:introduction}

The recent charged current measurement at the Sudbury Neutrino
Observatory (SNO)~\cite{Ahmad:2001an} has shed more light on the
long-standing problem posed by the previous solar neutrino
data~\cite{sun-exp} forcing us to reconsider the status of the
various solutions to the solar neutrino anomaly.

In this paper we re-consider the case of spin-flavour precession
solutions, based on non-zero transition magnetic moments of neutrinos
~\cite{Schechter:1981hw}.  These are especially attractive for several
reasons: (i) on general theoretical grounds \cite{Schechter:1980gr}
neutrinos are expected to be Majorana particles; (ii) such conversions
induced by transition magnetic moments may be resonantly amplified in
the Sun~\cite{Akhmedov:1988uk}; (iii) they offer the best pre-SNO
global fit of solar neutrino data~\cite{Miranda:2001bi}, and (iv) an
SFP type solution, being an active--to--active conversion mechanism,
has the right features to reconcile the SNO CC and SuperKamiokande
results.

Finally, such solutions are rather robust if the arbitrariness in the
choice of the magnetic field profile in the solar convective
zone~\cite{Akhmedov:2000fj,Guzzo:1999sb,Derkaoui:2001wx} is removed in
a self-consistent way from magneto-hydrodynamics
theory~\cite{Miranda:2001bi}.

By generalizing our previous work~\cite{Miranda:2001bi} to the case of
non-zero neutrino mixing we obtain two new and important results: (i)
we recover the resonant small-mixing solution to the solar neutrino
problem found previously~\cite{Miranda:2001bi} and analyse its status
in the light of the new SNO and 1258--day SK results, and (ii) we find
a genuinely new non-resonant SFP solution in the so-called dark-side
of the neutrino mixing parameter~\cite{Fogli:1996ne,deGouvea:2000cq}.
Following \cite{Bahcall:2001zu} we choose to determine the allowed
solutions by considering only the total rates of the solar neutrino
experiments, ignoring first all the data from the Super-Kamiokande
measurements of the spectral energy distribution and the day-night
variations.  We find that these solutions, both the resonant spin
flavour precession solution (which we call RSFP) as well as a new
non-resonant one (NRSFP solution), provide excellent descriptions of
the solar rates, including the recent SNO CC result.  Subsequently we
demonstrate how these solutions predict a substantially flat recoil
energy spectrum of solar neutrinos in agreement with the observations
of the Super-Kamiokande experiment~\cite{sun-exp}.  Moreover, our
solutions are consistent with the non-observation of electron
anti--neutrinos from the sun~\cite{Barbieri:1991ed,Vogel:1999zy} in
the results of the LSD experiment~\cite{Aglietta:1996zu} as well as
SuperKamiokande~\cite{smy}.

This paper is organized as follows. In
section~\ref{sec:neutr-conv-prob} we discuss the neutrino evolution
and conversion probabilities, in section~\ref{sec:calculational} we
summarize the calculational and fit procedures we adopt, while we
summarize our results in section~\ref{sec:summary-discussion}.

\section{Neutrino Evolution and Survival/Conversion Probabilities}
\label{sec:neutr-conv-prob}

Motivated by the results from reactor neutrino
experiments~\cite{Apollonio:1999ae} and to some extent also from
atmospheric neutrinos~\cite{Gonzalez-Garcia:2001sq} we adopt, for
simplicity, a two-flavour RSFP scenario.  The Majorana neutrino
evolution Hamiltonian in a magnetic field in this case is well--known
to be four--dimensional~\cite{Schechter:1981hw},
%
\begin{equation}
i\left(
\begin{array}{l}
\dot{\nu}_{eL} \\
\dot{\bar{\nu}}_{eR} \\
\dot{\nu}_{\mu L} \\
\dot{\bar{\nu}}_{\mu R}
\end{array}
\right) = \left(
\begin{array}{cccc}
V_e -c_2\delta & 0 & s_2\delta & \mu B_+(t) \\
0 & - V_e - c_2\delta & - \mu B_-(t) & s_2\delta \\
s_2\delta & - \mu B_+(t) & V_{\mu} + c_2\delta & 0 \\
\mu B_-(t) & s_2\delta & 0 & - V_{\mu} + c_2\delta
\end{array}
\right) \left(
\begin{array}{c}
\nu_{eL} \\
\bar{\nu}_{eR} \\
\nu_{\mu L} \\
\bar{\nu}_{\mu R}
\end{array}
\right)~,  \label{master}
\end{equation}
where $c_2 = \cos 2\theta$, $s_2 = \sin 2\theta$, $\delta = \Delta
m^2/4E$, assumed to be always positive, are the neutrino oscillation
parameters; $\mu$ is the neutrino transition magnetic moment; $B_{\pm}
= B_x \pm iB_y$, are the magnetic field components which are
perpendicular to the neutrino trajectory; $V_e(t) = G_F\sqrt{2}(N_e(t)
- N_n(t)/2)$ and $V_{\mu}(t) = G_F\sqrt{2}(-N_n(t)/2)$ are the
neutrino vector potentials for $\nu_{eL}$ and $\nu_{\mu L}$ in the
Sun, given by the number densities of the electrons ($N_e(t)$) and
neutrons ($N_n(t)$).  When $\theta \to 0$ we recover the case treated
in~\cite{Miranda:2001bi} while as $B \to 0$ we recover the pure OSC
case. In our calculations of $P_i$ we use the electron and neutron
number densities from the BP00 model \cite{BP00} with the magnetic
field profile obtained in ref.~ \cite{Miranda:2001bi} for k=6 and $R_0
= 0.6 R_\odot$. We assume a transition magnetic moment of $10^{-11}$
Bohr magneton, consistent with experiment and a magnetic field
magnitude around 80 kGauss, allowed by helioseismological
observations. Finally, in order to obtain Earth matter effects we
integrate numerically the evolution equation in the Earth matter using
the Earth density profile given in the Preliminary Reference Earth
Model (PREM) \cite{PREM}.

\subsection{The  solar neutrino conversion probability}
\label{subsec:prob}

The combined amplitude for a solar $\nu_e$ to be detected as $\nu_\alpha$ ( $\alpha$
being e, $\mu$, $\bar{e}$, $\bar{\mu}$) with energy $E$ at a detector in
the Earth can be written as:
\begin{equation}
A^{\text{S-V-E}}_{\nu_e\to\nu_\alpha}
=
\langle \nu_\alpha | U^{Earth}U^{Vacuum}U^{Sun}| \nu_e \rangle
=
\sum_{i=1,2,\bar{1},\bar{2}}A^S_{e\,i}\,A^E_{i\,\alpha}\,\exp[-im_i^2
(L-R_\odot)/2E]~
\,.
\label{amplitud}
\end{equation}

Here $A^S_{e\,i}$ is the amplitude of the transition $\nu_e \to \nu_i$
($\nu_i$ is the $i$-mass eigenstate) from the production point to the
Sun surface, $A^E_{i\,\alpha}$ is the amplitude of the transition $\nu_i \to
\nu_\alpha$ from the Earth surface to the detector, and the propagation in
vacuum from the Sun to the surface of the Earth is given by the
exponential, where $L$ is the distance between the center of the Sun
and the surface of the Earth, and $R_\odot$ is the radius of the Sun.
While the presence of magnetic field couples the four states in the
evolution, its absence in vacuum and in the Earth produces the
decoupling of the four states into two doublets : ( $\nu_e$, $\nu_{\mu}$ )
and ( $\nu_{\bar{e}}$, $\nu_{\bar{\mu}}$ ).  The corresponding probabilities
$P_{e \alpha}$ are then given by:
\begin{eqnarray}
P_{ee}&=&P_1P_{1e}+P_2P_{2e}+2\sqrt{P_1P_2P_{1e}P_{2e}}\cos\xi_1\\
\label{Pee}
P_{e\mu}&=&P_1P_{1\mu}+P_2P_{2\mu}-2\sqrt{P_1P_2P_{1\mu}P_{2\mu}}\cos\xi_1\\
\label{Pemu}
P_{e\bar{e}}&=&P_{\bar{1}}P_{\bar{1}\bar{e}}+P_{\bar{2}}P_{\bar{2}\bar{e}}-2\sqrt{P_{\bar{1}}P_{\bar{2}}P_{\bar{1}\bar{e}}P_{\bar{2}\bar{e}}}\cos\xi_2 ,\\
\label{Peae}
P_{e\bar{\mu}}&=&P_{\bar{1}}P_{\bar{1}\bar{\mu}}+P_{\bar{2}}P_{\bar{2}\bar{\mu}}+2\sqrt{P_{\bar{1}}P_{\bar{2}}P_{\bar{1}\bar{\mu}}P_{\bar{2}\bar{\mu}}}\cos\xi_2 .
\label{Peamu}
\end{eqnarray}
Here $P_i\equiv |A^S_{e\,i}|^2$ is the probability that the solar neutrinos
reach the surface of the Sun as $|\nu_i\rangle$, while $P_{i \alpha} \equiv
|A^E_{i\,\alpha}|^2$ is the probability of $\nu_i$ arriving at the surface of
the Earth to be detected as $\nu_\alpha$.  The phases $\xi_a$ ($a=1,2$) are
given by
\begin{eqnarray}
\xi_a=\frac{\Delta m^2 (L-R_\odot)}{2E}+\phi_a\, ,
\label{int}
\end{eqnarray}
where $\phi_a$ contain the phases due to propagation in the Sun
and in the Earth and we checked that it can be safely neglected for
our purposes.

\begin{figure}[t] \centering
    \includegraphics[height=14cm,angle=-90]{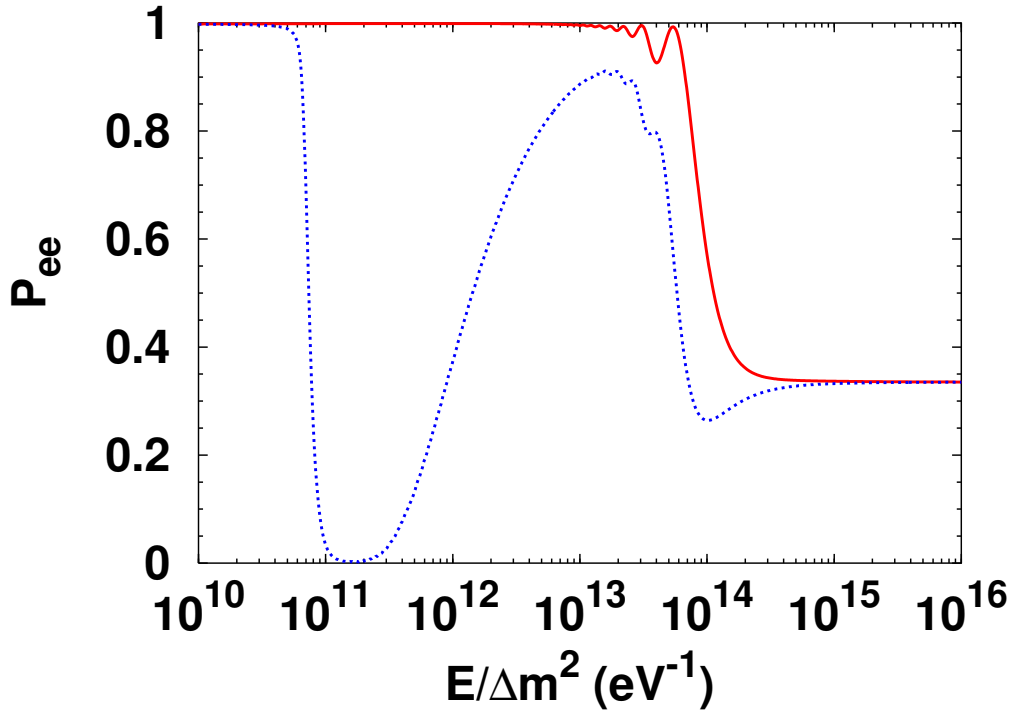} \hfil

    \caption{\label{fig:prob} %
      Neutrino spin flavour precession survival probabilities in
      ``light'' and ``dark'' sides, for $\mu = 10^{-11} \mu_B$ and
      $B_\perp \sim 80$ kGauss.}
\end{figure}

The results presented in the following sections have been obtained
using the general expression for the probabilities with
$P_1$,$P_2$,$P_{\bar{1}}$ and $P_{2e}$ found by numerically solving
the evolution equation~(\ref{master}). The probabilities required in
Eq.~(\ref{Pee}) are not independent from the last ones and can be
obtained using the unitarity relations and using relations between
both octants in mixing for the evolution in the Earth, ie.,
$P_{\bar{1}\bar{e}}(\theta)=P_{2e}(\frac{\pi}{2} - \theta)$. In the
limit $B_\perp \to 0$ we recover the forms given in
ref.~\cite{Gonzalez-Garcia:2000sk}.  In Fig.~\ref{fig:prob} we show a
schematic view of the spin flavour precession survival probabilties
both in the ``light'' and ``dark'' sides. The first corresponds to $0
\leq \theta \leq \pi/4$ while the latter means $ \pi/4 \leq \theta
\leq \pi/2$.  The dotted curve corresponds to the RSFP case, while the
solid one will lead to the new non-resonant NRSFP solution, see below.
One notices that, in contrast to the oscillation case, the asymptotic
value of the survival probability in the SFP model can be lower than
0.5 as $E \to \infty $ or $\Delta m \to 0$. This can be understood as follows.
Consider the idealized case of constant matter potential, constant
magnetic field over a finite slab $\Delta r$ and $\cos2\theta \approx \pm 1$.  In this
case one can write simple analytic formulae for the neutrino
conversion probabilities. For example the $\nu_e$ survival probability
may be given as
  \begin{multline}
 P_{ee} =1- \frac{(2\mu_\nu B_{\perp})^2}{(V_e+V_\mu - 2 \delta
\cos 2\theta)^2 + (2\mu_\nu B_{\perp})^2}\times\\
 \times\sin^2
\left(\sqrt{(V_e+V_\mu- 2 \delta \cos
2\theta)^2 + (2\mu_\nu B_{\perp})^2}\frac{\Delta r}{2}\right)~.
  \label{eq:sfprob}
\end{multline}
For the case $\cos2\theta \approx 1$ we obtain the well-known resonant solution,
while the alternative $\cos2\theta \approx -1$ choice corresponds to our new
NRSFP solution in the dark side, see below. The higher asymptotic
suppresion of $P_{ee}$ in both cases implies a higher possible degree
of suppression of $^8$ B neutrinos than achievable in the OSC case.
Moreover the converted $\bar\nu_\mu$ can be detected via the neutral
current, thus reconciling the SNO CC result with the higher
Super-Kamiokande rate measurement.

\section{Calculational Method}
\label{sec:calculational}

In our following description of solar neutrino data~\cite{sun-exp} we
adopt the analysis techniques which have already been presented in
previous papers~\cite{Bahcall:2001zu,Gonzalez-Garcia:2001sq,two} using
the theoretical BP00 standard solar model best--fit fluxes and
estimated uncertainties from ref.~\cite{BP00}.  In addition to the
solar data~\cite{sun-exp} we also use the reactor
data~\cite{Apollonio:1999ae} as well as the data on searches for
anti-neutrinos from the sun~\cite{Aglietta:1996zu}.  For the neutrino
conversion probabilities we use the numerical results calculated in
the previous section.

We employ the self-consistent magneto-hydrodynamics magnetic field
profile obtained in ref.~ \cite{Miranda:2001bi} for k=6 and $R_0 = 0.6
R_\odot$.  The resulting theoretical framework has therefore only 2
effective free parameters: $\Delta m^2$, $\tan^2\theta$. The remaining
parameter $\mu B_\perp$ characterizing the maximum magnitude of the
magnetic field in the convective zone has been fixed at its optimum
value.  Since the parameter space is three-dimensional, the allowed
regions for a given C.L. are defined as the set of points satisfying
the condition
\begin{equation}
    \chi^2_{\rm SOL}(\Delta m^2,\theta, \mu B_\perp  )
    -\chi^2_{\rm SOL,min}\leq \Delta\chi^2 \mbox{(C.L., 2~d.o.f.)} ,
\label{eqn:chirsf}
\end{equation}
where $\chi^2_{\rm SOL}$ contains
\begin{equation}
\chi^2_{LSD}(\Delta m^2, \theta, \mu B_{\perp}) =
        \frac{\left(N^{TH}_{\bar{\nu_e}}(\Delta m^2, \theta, \mu B_{\perp}) - N^{EXP}_{\bar{\nu_e}}\right)^2}
                        {\sigma_{LSD}^2}
\end{equation}
where $N^{EXP}_{\bar{\nu}} = - 1.5$ and $\sigma_{LSD}=22$ in order to
account for the data on searches for anti-neutrinos from the
sun~\cite{Aglietta:1996zu}. As we will see this term plays an
important role in restricting the neutrino parameters.

In our numerical calculations we use the survival/conversion
probabilities of solar electron neutrino valid in the full range of
$\Delta m^2$ and $\theta$, selecting the optimum value of $\mu
B_\perp$ with $B_\perp$ varying over the range from 0 to 100
kGauss~\footnote{A description of this procedure will be presented
  elsewhere~\cite{inprep}.}.

Finally, we employ the relevant reaction cross sections and
efficiencies for the all experiments used in
ref.~\cite{Bahcall:2001zu,Gonzalez-Garcia:2001sq,two}. For the SNO
case the CC cross section for deuterium was taken from~\cite{crossno}.

\subsection{Rate Fit }
\label{sec:rate-fit-}

Here we take into account the total rates in the chlorine, gallium,
and Super-Kamiokande experiments, the SNO CC result and the
anti-neutrino limit from LSD, and also the reactor neutrino
data~\cite{Apollonio:1999ae}. The rates from the GALLEX/GNO
experiments have been averaged so as to provide a unique data point.
The resulting number of degrees of freedom is therefore 4: 4 (rates) +
SNO + LSD $-$2 (parameters: $\Delta {\rm m}^2$, $\theta$) with a fixed
$\mu B_\perp$.

\begin{figure}[t] \centering
    \includegraphics[height=12cm,angle=0]{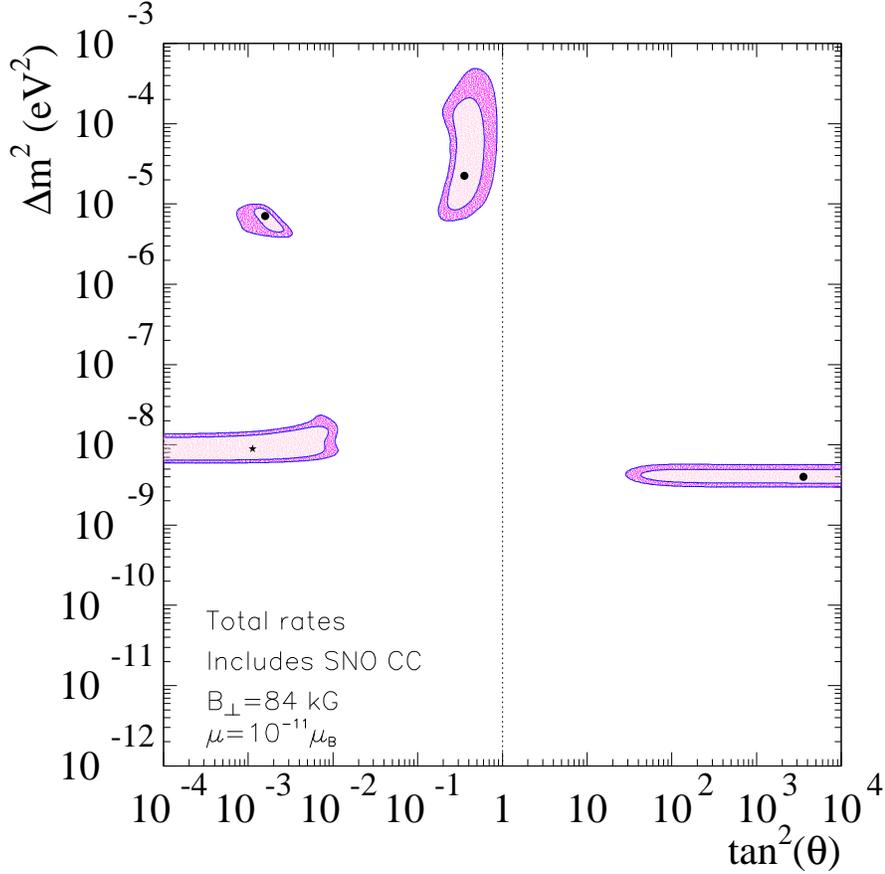} \hfil

    \caption{\label{fig:ratesonly} %
      Allowed solutions to the solar neutrino rates and reactor data
      for $\mu_\nu = 10^{-11} \mu_B$ and $B_\perp = 84$ kGauss.  The
      upper limit on the solar anti-neutrino flux according to LSD
      data is included.}
\end{figure}

We present in Fig.~\ref{fig:ratesonly} the allowed solutions for the
two-flavour SFP case. These include the pure two-neutrino oscillation
case, as well as the convenitional RSFP and the new NRSFP solution.

Note that the contours refer to 90\%, 99\% CL defined with respect to
the global minimum of $\chi^2$.  We find that both LMA and SMA
oscillation solutions are recovered without an essential change due to
the effect of the magnetic moment.  The SMA solution appears (even
though disfavored), but leads to an unacceptably tilted recoil energy
spectrum, as will be seen in Fig.~\ref{fig:spec}.

An important point to notice is that this plot lacks the LOW solution
as well as the characteristic region joining it through the dark side
to the vacuum-type solutions~\cite{inprep}. In this figure we have
adjusted the value of $\mu B_\perp$ to its best value (for $\mu = 10^{-11}$
Bohr magneton this corresponds to $B_\perp \sim 80$ kGauss).  One sees that
the relatively large $\mu$ value has important consequences. It leads
in this case in the complete absence of all large mixing solutions
other than the LMA solution due to the effect magnetic field.  Such a
value implies an important modification in the neutrino survival
probability implying an unwanted over-suppression of the $^8$B
neutrino flux and therefore the impossibility to account for all
experiments in this region because of the high $\bar\nu_e$ flux.  From
this point of view vacuum-type solutions are \textsl{unstable} against
the effect of the magnetic field. In fact the non-LMA large mixing OSC
solutions are not re-instated even if the $^8$B neutrino flux is left
free.  The goodness of fit of the various solutions in
Fig.~\ref{fig:ratesonly} is given in table~1.  One notices that, of
the OSC-type solutions, LMA is the best~\footnote{The first time the
  LMA solution was shown to be the best OSC solution was in
  ~\cite{two} due to the details of the solar neutrino spectra
  measured at Super-Kamiokande.  This trend is now re-inforced by the
  enhanced statistics. The SNO CC rate-result implies, on its own, a
  preference for the LMA if the BP00 boron flux is assumed}.  However
the SFP solutions are slightly better.

\begin{table}[htbp]
\begin{center}
\begin{tabular}{lcccc}
\noalign{\bigskip}
\hline
\noalign{\smallskip}
Solution&$\Delta m^2$&$\tan^2(\theta)$& $\chi^2_{\rm min}$ &g.o.f. \\
\noalign{\smallskip}
\hline
\noalign{\smallskip}
LMA & $2.1 \times 10^{-5} $  &$ 0.34 $ & 3.99 &$14$\% \\
SMA & $ 6.9 \times 10^{-6} $  &  $1.6  \times 10^{-3} $ &  5.25  &$7$\%\\
RSF & $8.9 \times 10^{-9}$  &  $1.1  \times 10^{-3} $ & 2.98 &$22$\%\\
NRSF & $4.0 \times 10^{-9} $  & $3.5  \times 10^{3}$ &  3.83  &$15$\%\\
\noalign{\smallskip}
\hline
\end{tabular}
 \caption{ Best-fit points and goodness-of-fit of  oscillation
   and spin flavour solutions to the solar neutrino problem as
   determined from the rates-only analysis for $\mu = 10^{-11}
   \:\mu_B$ and $B_\perp = 84$ kGauss.}
  \end{center}
\end{table}

Note that the goodness-of-fit given in the last column is calculated
using the value of $\chi^2/{\rm d.o.f}$ for each allowed solution
corresponding to each of the local minima of table~1.  Note also that
should we perform a restricted two-parameter analysis using only $\Delta
m^2$ and the neutrino mixing angle $\theta$ for the pure OSC case and only
$\Delta m^2$ and $\mu B_\perp$ for the pure SFP case we obtain exactly the
same goodness-of-fit and $\chi^2_{MIN}$ for each of the corresponding
SFP and OSC solutions in table~1.

A more striking feature of Fig.~\ref{fig:ratesonly} is the appearance
of two new solutions which are totally due to the effect of the
magnetic field. One contains the previous resonant no-mixing solution
which is recovered, after updating the solar data to the measurements
from 1258 days of Super-Kamiokande data and SNO CC measurement. One
sees that this RSFP solution extends up to $\tan^2\theta$ values around
$10^{-2}$ or so.  More importantly, one finds a genuinely new
non-resonant (NRSFP) solution in the ``dark-side'' of the parameter
space, for large $\tan^2\theta$ values.  The existence of these solutions
can be easily understood on the basis of Fig.~\ref{fig:prob}.
Similarly one can understand the non-resonant nature of the new NRSFP
solution.
Note that in obtaining the shape of the RSF solutions we have made use
of the data on searches for anti-neutrinos from the
sun~\cite{Aglietta:1996zu}. These play an important role in cutting
the non-resonant RSF solution to $\tan^2\theta$ values larger than
about 30.

\subsection{Recoil Spectra}
\label{sec:recoil-spectra}

We now present the predicted day--night averaged~\footnote{Note that,
  in contrast to the OSC spectra, where a day-night effect is
  predicted, the SFP spectra show no day-night asymmetry.}  spectral
energy distribution for our two spin flavour precession solutions and
compare it with those of the pure OSC-type solutions.

\begin{figure}[t] \centering
    \includegraphics[height=12cm,angle=0]{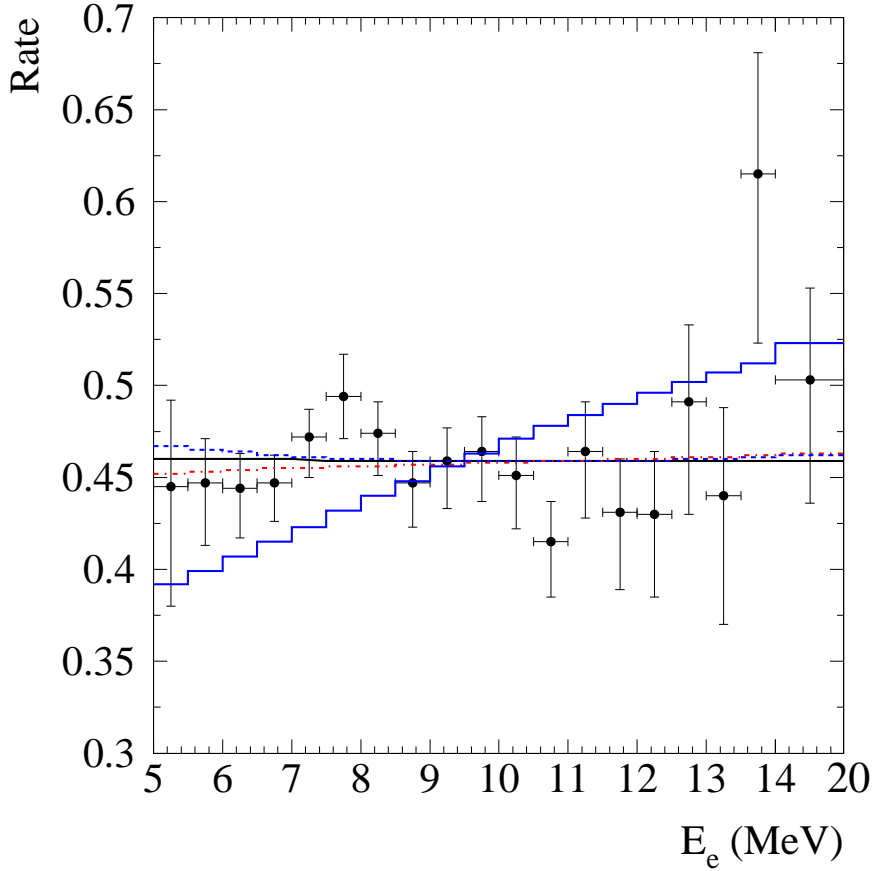} \hfil

    \caption{\label{fig:spec} %
      Predicted recoil energy spectra for spin flavour precession
      solutions. Details in text.}
\end{figure}
In Fig.~\ref{fig:spec} we present the recoil energy spectra for spin
flavour precession. The thin solid horizontal line corresponds to the
new NRSFP solution, while the dot-dashed refers to the standard RSFP
solution. Clearly both spin flavour precession spectra are totally
consistent with the Super-Kamiokande data and, as a result, will
remain as excellent solutions after the inclusion of the recoil energy
spectra.  We also present the predicted oscillation spectra, in solid
that of the SMA solution and dashed the LMA solution. Clearly one can
see that, in contrast with the RSFP and NRSFP solutions, the SMA
spectrum is in strong disagreement with the SK data.  A full-fledged
global fit of the recoil spectra for the spin flavour solutions is
outside the scope of this letter and will be presented
elsewhere~\cite{inprep}.

\section{Summary and Discussion}
\label{sec:summary-discussion}

In this paper we have re-considered the case of spin-flavour
precession solutions, based on non-zero transition magnetic moments of
Majorana neutrinos taking into account the recent SNO CC result as
well as the 1258--day solar neutrino data from Super-Kamiokande. We
have also payed attention to the upper limit on the solar
anti-neutrino flux from the LSD experiment as well as the reactor
neutrino data. We have followed the self-consistent approach from
magneto-hydrodynamics theory employed previously~\cite{Miranda:2001bi}
in order to remove the arbitrariness associated to the magnetic field
profile~\cite{Akhmedov:2000fj,Guzzo:1999sb,Derkaoui:2001wx}. This
effectively reduces the theoretical analysis framework to a
three-parameter one. In-so-doing we have also generalized our previous
work~\cite{Miranda:2001bi} to the case of non-zero neutrino mixing,
performing the first ``unified'' study of solar neutrino data in the
presence of a neutrino transition magnetic moment. It contains as
particular cases ``light-side'' and ``dark-side'' OSC as well as
genuine SFP solutions.

We have recovered the standard resonant small-mixing solution to the
solar neutrino problem (RSFP) which remains as best solution to the
solar neutrino anomaly (see table). Second, we have found a genuinely
new non-resonant solution in the so-called ``dark-side'' of the
neutrino mixing parameter. Such NRSFP solution gives a very good fit
of the present solar neutrino data.  Although we have chosen to
determine the allowed solutions by considering only the total rates of
the solar neutrino experiments, we have presented their predicted
recoil spectra, showing how they are in agreement with the data from
the Super-Kamiokande experiment.  A full comparative study of
oscillation and spin flavour solutions of the solar neutrino problem
is outside the scope of this letter and will be presented
elsewhere~\cite{inprep}.


\section*{Acknowledgements}

We would like to thank Alexei Bykov, Hiroshi Nunokawa, Alexander Rez,
Victor Popov and Dmitri Sokoloff for very useful discussions.  This
work was supported by Spanish grants PB98-0693 and GV99-3-1-01, by the
European Commission RTN network HPRN-CT-2000-00148 and by the European
Science Foundation network grant N.~86.  VBS and TIR were partially
supported by the RFBR grants 00-02-16271 and 01-02-06225 and OGM was
supported by the CONACyT-Mexico grants J32220-E and 35792-E. TIR and
OGM thank the Val{\`e}ncia Astroparticle and High Energy Physics Group for
the kind hospitality.


\end{document}